%% file: main.tex
\input{settings}
\usepackage{graphicx}
\usepackage[caption=false]{subfig}
\usepackage{orcidlink}
\definecolor{DH-color}{named}{orange}
\definecolor{DH-color2}{named}{red}

\definecolor{PP-color}{named}{red}
\definecolor{PP-color2}{named}{red}

\definecolor{PS-color}{named}{black}
\definecolor{PS-color2}{named}{red}

\begin{document}
\title{Lindblad equation for the quantum Rabi model in the deep strong coupling regime}
\author{Daniel Holst\,\orcidlink{0009-0007-4875-2454}}
\email{daniel.holst@fysik.lu.se}
\affiliation{Physics Department and NanoLund, Lund University, Box 118, 22100 Lund, Sweden.}
\author{Patrick Potts\, \orcidlink{0000-0001-6036-7291}}
\affiliation{Department of Physics and Swiss Nanoscience Institute,
University of Basel, Klingelbergstrasse 82, 4056 Basel, Switzerland}
\author{Peter Samuelsson\,\orcidlink{0000-0001-5025-4832}}
\affiliation{Physics Department and NanoLund, Lund University, Box 118, 22100 Lund, Sweden.}

\date{\today}

\begin{abstract}
    We derive a quantum master equation in Lindblad form for the open quantum Rabi model in the perturbative deep strong coupling regime, where the spectrum is close to degenerate and harmonic. The Lindblad form is obtained by combining secular and singular-coupling approximations, working in a Schrieffer-Wolff transformed frame. We benchmark our equation against a Bloch-Redfield one, analyzing the time-evolution and steady state properties of several observables and find excellent agreement. Analytical results are obtained in limiting cases.

\end{abstract}

\maketitle

\input{intro_ps}

\input{equation_ps}

\input{results_ps}

\textit{Conclusions}---We have derived a compact Lindblad equation for the dynamics of the quantum Rabi model in the pDSC regime. Benchmarking against a Bloch-Redfield equation we find excellent agreement. Analytical expressions are derived in limiting cases, giving an improved qualitative understanding of the pDSC regime. Our results will support further experiments on systems described by the quantum Rabi model in the DSC regime, with potential for novel application in quantum technology.
\\
\\
\textit{Acknowledgments}---DH thanks Anton Frisk Kockum, Gernot Schaller, Lei Du and Sofia Sevitz for fruitful discussions. P.S. and D.H. were supported by the Swedish Research Council, Grant No. 2023-03915. P.P.P. acknowledges funding from the Swiss National Science Foundation, Eccellenza Professorial Fellowship PCEFP2\textunderscore194268. P.S acknowledges support by the Wallenberg Centre for Quantum Technology (WACQT) funded by Knut and Alice Wallenberg (KAW) Foundation. We acknowledge limited use of AI tools ChatGPT 5.6 Sol and Claude Opus 5 for input on analytical and numerical calculations. All ideas originated among the authors who take full responsibility for the results and conclusions presented. 
\\
\\

\bibliographystyle{apsrev4-1}
\bibliography{biblography}

\newpage
\begin{widetext}
\section*{End Matter}    
\end{widetext}

\input{endmatter_ps}

\newpage
\begin{widetext}

\section*{Supplemental Material}
\setcounter{equation}{0}
\renewcommand{\theequation}{S\arabic{equation}}
Throughout the Supplemental Material, standard equation numbers refer to equations in the main text and equation numbers starting with an S refer to equations in the Supplemental material.

\input{appendix_final_ps}
\end{widetext}
\end{document}

%% file: settings.tex
\documentclass[%
superscriptaddress,
twocolumn,
amsmath,amssymb,
aps,prl,
,10pt]{revtex4-2}

\usepackage{graphicx}
\usepackage{dcolumn}
\usepackage{bm}
\usepackage[english]{babel}

\usepackage[numbered, framed]{matlab-prettifier}
\def\shrug{\texttt{\raisebox{0.75em}{\char`\_}\char`\\\char`\_\kern-0.5ex(\kern-0.25ex\raisebox{0.25ex}{\rotatebox{45}{\raisebox{-.75ex}"\kern-1.5ex\rotatebox{-90})}}\kern-0.5ex)\kern-0.5ex\char`\_/\raisebox{0.75em}{\char`\_}}}

\DeclareFontFamily{OMX}{MnSymbolE}{}
\DeclareSymbolFont{MnLargeSymbols}{OMX}{MnSymbolE}{m}{n}
\SetSymbolFont{MnLargeSymbols}{bold}{OMX}{MnSymbolE}{b}{n}
\DeclareFontShape{OMX}{MnSymbolE}{m}{n}{
    <-6>  MnSymbolE5
   <6-7>  MnSymbolE6
   <7-8>  MnSymbolE7
   <8-9>  MnSymbolE8
   <9-10> MnSymbolE9
  <10-12> MnSymbolE10
  <12->   MnSymbolE12
}{}
\DeclareFontShape{OMX}{MnSymbolE}{b}{n}{
    <-6>  MnSymbolE-Bold5
   <6-7>  MnSymbolE-Bold6
   <7-8>  MnSymbolE-Bold7
   <8-9>  MnSymbolE-Bold8
   <9-10> MnSymbolE-Bold9
  <10-12> MnSymbolE-Bold10
  <12->   MnSymbolE-Bold12
}{}

\newcommand{\cre}{\hat{a}^\dagger}
\newcommand{\des}{\hat{a}}

\usepackage{svg}
\usepackage{braket}
\definecolor{ForestGreen}{RGB}{34,139,34}

\newcommand{\ham}{\hat{H}}

\usepackage{xcolor}
\definecolor{amber}{rgb}{1.0, 0.49, 0.0}
\definecolor{cadmiumgreen}{rgb}{0.0, 0.42, 0.24}

\usepackage{amsthm}
\usepackage{tikz}

\newtheoremstyle{styleth}%
{3pt}
{3pt}
{}
{}
{\bfseries\color{amber}}
{}
{.5em}
{}
\theoremstyle{styleth}

\newtheoremstyle{styledef}%
{3pt}
{3pt}
{}
{}
{\bfseries\color{cadmiumgreen}}
{}
{.5em}
{}
\theoremstyle{styledef}

\usepackage{amsmath}

\usepackage{ragged2e}
\usepackage{blindtext}

\definecolor{ForestGreen}{RGB}{34,139,34}

\newcommand{\crec}{\hat{c}^\dagger}
\newcommand{\desc}{\hat{c}}
\usepackage{enumitem,xcolor}
\usepackage{ulem}

\usepackage[section]{placeins}


%% file: intro_ps.tex
\textit{Introduction}---The quantum Rabi model, describing a two-level system - a qubit -  coupled to a single electric field mode - an oscillator - constitutes a fundamental quantum model of light-matter interaction \cite{xie_quantum_2017}. Almost two decades ago, experiments with semiconductor polaritons \cite{anappara_signatures_2009,Scalari2012} and superconducting circuits \cite{niemczyk_circuit_2010,forn-diaz_observation_2010} reached qubit-oscillator couplings strong enough for the rotating wave approximation, reducing the quantum Rabi model to the Jaynes-Cummings one \cite{ShoreKnight1993}, to break down. In this ultrastrong coupling (USC) regime \cite{friskkockumUltrastrongCouplingLight2019,forn-diaz_ultrastrong_2019}, where the full Rabi model is needed to describe the physics, the ground state is entangled \cite{AshhabNori2010}, displaying virtual photon excitations \cite{Cirio2016,Gunter2009,DeLiberato2017}. Moreover, well known phenomena such as the Purcell effect, are suppressed \cite{de_liberato_light-matter_2014}, and the cavity photon emission statistics is altered \cite{Ridolfo2013}. More recently, even larger couplings, of the order of the oscillator and qubit frequencies and beyond, were reached. This deep strong coupling (DSC) regime \cite{casanovaDeepStrongCoupling2010a} was demonstrated intrinsically, in platforms based on circuit QED \cite{yoshihara_superconducting_2017,Yoshihara2017PRA,yoshihara_inversion_2018,Wang2023}, Landau polaritons \cite{bayer_terahertz_2017,Mornhinweg2024} and plasmonic nanoparticle crystals \cite{mueller_deep_2020} and also quantum simulated with circuit QED \cite{Langford2017}, trapped ions \cite{Lv2018,Cai2021} and cold atoms \cite{koch_quantum_2023,Hunanyan2024}.  

 \begin{figure}[t]
    \includegraphics[width=0.95\linewidth]{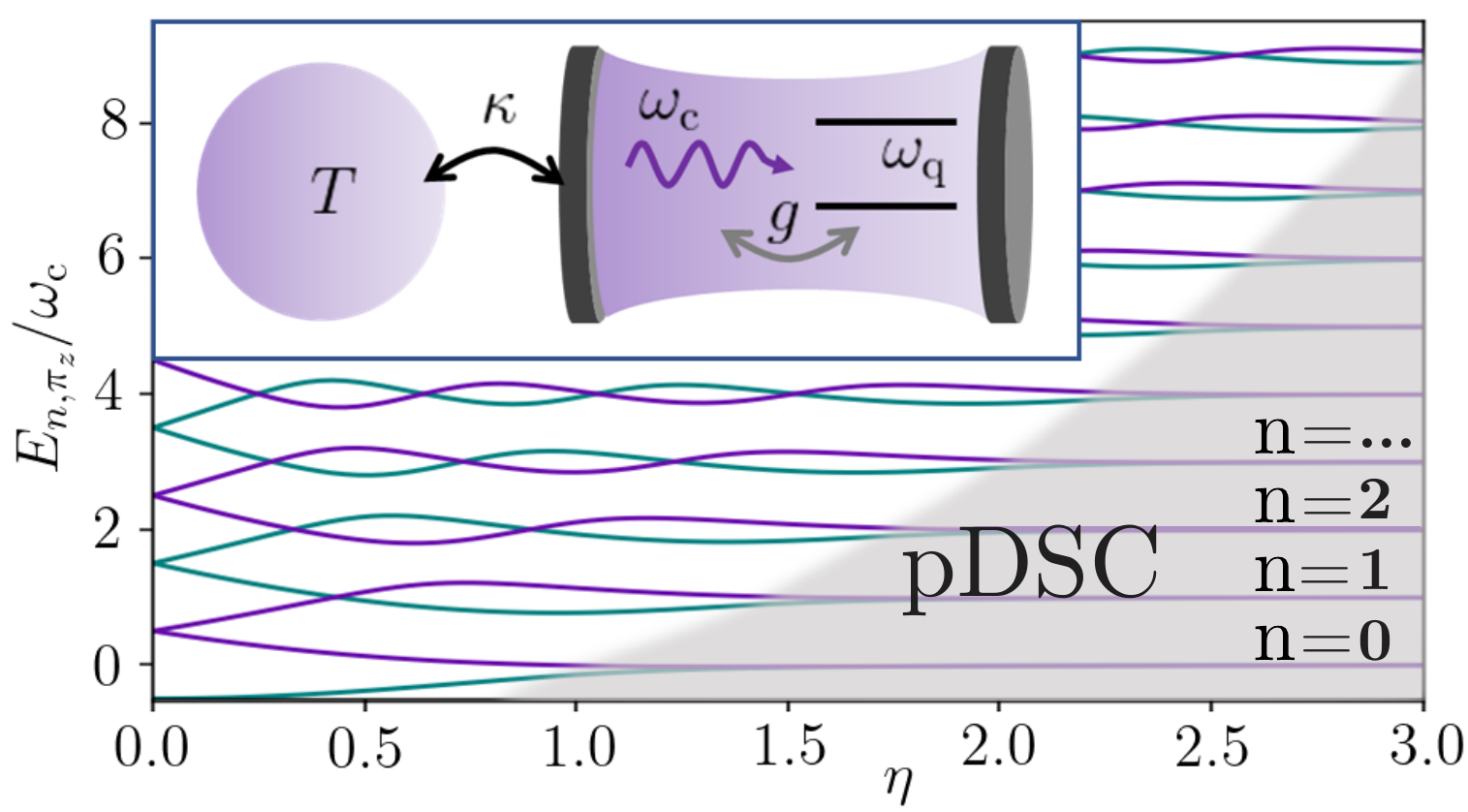}
    \caption{Lowest energy eigenvalues $E_{n,\pi_z}$ of the quantum Rabi Hamiltonian as a function of dimensionless coupling $\eta=g/\omega_\text c$ for $\omega_\text q=\omega_\text c$. Teal/Purple lines correspond to states with $+/-$ parity. The shaded area shows the perturbative deep strong coupling (pDSC) regime. Inset: Schematic of the system with a qubit, frequency $\omega_\text q$, coupled with strength $g$ to a single cavity mode, frequency $\omega_\text c$. The cavity mode is further coupled, with strength $\kappa$, to a bath kept at temperature $T$.}
    \label{fig:system}
\end{figure} 

For open quantum Rabi systems (see inset in Fig. \ref{fig:system}) in the USC and DSC regimes it was shown \cite{beaudoinDissipationUltrastrongCoupling2011} that a quantum master equation, describing the dynamics of the Rabi system's reduced density matrix, should be formulated in terms of the eigenstates of the joint qubit-oscillator system to avoid unphysical results. This dressed state approach was generalized \cite{settineriDissipationThermalNoise2018} to systems with arbitrary spectrum, including harmonic and quasi-harmonic ones, resulting in a quantum master equation not in Lindblad form and hence not guaranteeing positivity of the density matrix. Further works have derived equations in Lindblad form in the adiabatic regime, with the qubit frequency much smaller than the oscillator one \cite{CastanosCervantes2022}, or qubit-cavity de-tuned regime \cite{Bina2012}. Methods where the environment is replaced by a set of lossy, interacting modes \cite{Lednev2024}, have also been employed. For large parts of the USC and DSC regimes there are however no simple, analytically tractable Lindblad equations. This paucity might prevent a better understanding of the open Rabi dynamics in these regimes and hamper possible quantum technology applications \cite{Romero2012,Nataf2011} based thereon.

Here we contribute to filling this gap by deriving a Lindblad equation valid in the perturbative DSC (pDSC) regime \cite{rossattoSpectralClassificationCoupling2017c}, where the quantum Rabi model simplifies to describe two weakly coupled, displaced oscillators with quasi-degenerate, harmonic spectra \cite{braakIntegrabilityRabiModel2011,Chen2012}, see Fig. \ref{fig:system}. We show that the dissipative dynamics for infinite coupling takes place by nearest-neighbor hopping along two uncoupled, total-parity conserved chains of displaced oscillator states. A finite coupling gives rise to a perturbative coupling of the chain dynamics. The Lindblad equation is derived by employing a unified approach \cite{trushechkinUnifiedGoriniKossakowskiLindbladSudarshanQuantum2021,pottsThermodynamicallyConsistentMarkovian2021a}, combining secular and singular-coupling approximations, together with a Schrieffer-Wolff transformation. Benchmarking against a Bloch-Redfield equation gives excellent agreement in the pDSC-regime and analytical results are presented in several limiting cases. This illustrates the accuracy, simplicity and usefulness of the equation for a more in-depth analysis of the pDSC regime.

%% file: equation_ps.tex
\textcolor{PS-color}{\textit{Derivation of Lindblad equation}--- We work in the dipole gauge, see Appendix A for details, where the Rabi Hamiltonian, describing a qubit with energy splitting $\omega_\text q$ coupled with strength $g$ to an oscillator - here a cavity mode for concreteness - with frequency $\omega_c$ (see Fig. \ref{fig:system}), is given by}
\begin{equation}
    \ham_\text{R} = \omega_\text c \cre\des + \frac{\omega_\text q}{2}\hat{\sigma}_z + g\hat{\sigma}_x(\cre+\des).
    \label{eq:Rabi}
\end{equation}
\textcolor{PS-color}{Here $\cre/\des$ are cavity photon creation/annihilation operators and $\hat \sigma_z$ and $\hat \sigma _x$ are qubit Pauli operators. The DSC regime is conveniently analyzed \cite{casanovaDeepStrongCoupling2010a} by introducing the excitation parity operator $\hat \pi_z=\hat \sigma_z(-1)^{(\hat a^\dagger\hat a+1)}$ and new ladder operators $\hat c,\hat c^\dagger$,  where $\desc = \hat{\sigma}_x\des$, noting that $[\hat c,\hat \pi_z]=0$. Further, by rotating to a displaced frame, $\ham_\text{D}=\hat D(\eta) \ham_\text{R} \hat D^\dagger(\eta)$ with $\hat{D}(\eta) = e^{\eta(\crec-\desc)}$ and $\eta=g/\omega_\text c$, the coupling term is removed from the Hamiltonian, giving (up to a constant term)}
\begin{equation}
    \ham_\text{D} = \omega_\text c \crec\desc - \frac{\omega_\text q}{2}\hat{\pi}_z (-1)^{\crec\desc}\hat{D}(-2\eta).
    \label{eq:dispH}
\end{equation}
\textcolor{PS-color}{Since $[\hat H_\text{D},\hat \pi_z]=0$, the Hamiltonian describes two harmonic oscillators (HO), one for each parity $\pi_z =\pm$, perturbed (with opposite sign) by the $\eta$ dependent term $(\omega_\text q/2)(-1)^{\crec\desc}\hat{D}(-2\eta)$. The eigenvalues $E_{n,\pi_z}$ of $\hat H_\text{D}$ are shown as a function of $\eta$ in Fig. (\ref{fig:system}) for the lowest energies, forming quasi-degenerate pairs, $n=0,1,2...$, with opposite parity in the DSC regime $\eta>1$.}

\textcolor{PS-color}{The pDSC regime, shown shaded in Fig. (\ref{fig:system}), is formally defined by $|E_{n,+}-E_{n,-}| \ll \omega_\text c$, i.e. the parity pair splitting is much smaller than the HO level spacing \cite{rossattoSpectralClassificationCoupling2017c}. The energy eigenvalues are well described \cite{SM}, for $\omega_\text q \lesssim \omega_\text c$, by first order in perturbation theory in $(\omega_\text q/2)(-1)^{\crec\desc}\hat{D}(-2\eta)$, as $E_{n,\pi_z} \approx \epsilon_{n,\pi_z}$, with}
\begin{equation}
    \epsilon_{n,\pi_z}=n\omega_\text c+ \pi_z\frac{\omega_\text q}{2}D_{nn}
\label{eq:energy}
\end{equation}
\textcolor{PS-color}{with corresponding eigenstates $|n,\pi_z\rangle \approx |n,\pi_z\rangle_{0}+\pi_z\omega_\text q/(2\omega_\text c)\sum_{m\neq n} D_{mn}/(n-m)|m,\pi_z\rangle_0$. Here $|n,\pi_z\rangle_{0}=|n\rangle_0\otimes |\pi_z\rangle$  where $|n\rangle_0$ are unperturbed HO number states, $|\pi_z\rangle$ are the parity eigenstates, $\hat \pi_z|\pi_z\rangle=\pi_z|\pi_z\rangle$ and the $D_{nn}$ matrix elements are given by $D_{nm}=(-1)^{n+1}~_0\langle n|\hat D(-2\eta) |m\rangle_0$.}

\textcolor{PS-color}{The open quantum Rabi system is shown in the inset in Fig. \ref{fig:system}, with the cavity weakly coupled, $\kappa \ll \omega_c$, to a bosonic bath kept at temperature $T$. In the displaced frame, the system component of the system-bath coupling Hamiltonian is $\hat A=(\hat c^\dagger+\hat c)\hat \pi_x$ (with $\hat \pi_x=\hat \sigma_x)$. Following the standard techniques \cite{breuerTheoryOpenQuantum2009} for deriving a quantum master equation within the Born-Markov approximation, neglecting for simplicity external sources of noise, we arrive at the Bloch-Redfield (BR) equation for the reduced system density operator $\hat \rho^\text I(t)$ in the interaction picture,}
\begin{eqnarray}
   \partial_t\hat \rho^{\text I}=\frac{1}{2}\sum_{\omega,\omega^\prime} e^{i(\omega^\prime-\omega)t} \gamma(\omega) \mathcal{\tilde D}[\hat A, \omega,\omega^\prime]\hat \rho^\text{I}+\mbox{h.c.}, 
   \label{BR}
\end{eqnarray}
\textcolor{PS-color}{where the sums run over distinct Bohr frequencies $\omega, \omega^\prime=E_{m,\tau_z}-E_{n,\pi_z}$ and $\mathcal{\tilde D}[\hat A, \omega,\omega^\prime]\hat \rho^\text I =\hat A(\omega)\hat \rho^{\text I} \hat A^\dagger(\omega^\prime) - \hat A^\dagger(\omega^\prime)\hat A(\omega)\rho^{\text I}$ with $\hat A(\omega)=\sum_{n,m,\pi_z,\tau_z}|n,\pi_z\rangle\langle n,\pi_z|\hat A|m,\tau_z\rangle\langle m,\tau_z|\delta_{\omega, E_{m,\tau_z}-E_{n,\pi_z}}$. The rate $\gamma(\omega)=\kappa([1+n_\text B(\omega)]\theta(\omega)+n_\text B(-\omega)\theta(-\omega))$, with $\theta(\omega)$ the step function and $n_\text B(\omega)=1/(\exp(\beta \omega)-1)$ the Bose distribution function with $\beta=1/(k_\text BT)$. Neglecting the Lamb shift in Eq. (\ref{BR}), ensures \cite{Lee2022} that the (Schr\"odinger picture) steady state is the Gibbs state $e^{-\beta \hat H_\text{D}}/\mbox{tr}(e^{-\beta \hat H_\text{D}})$.}

\begin{figure}[ht]
    \centering
    \includegraphics[width=0.99\linewidth]{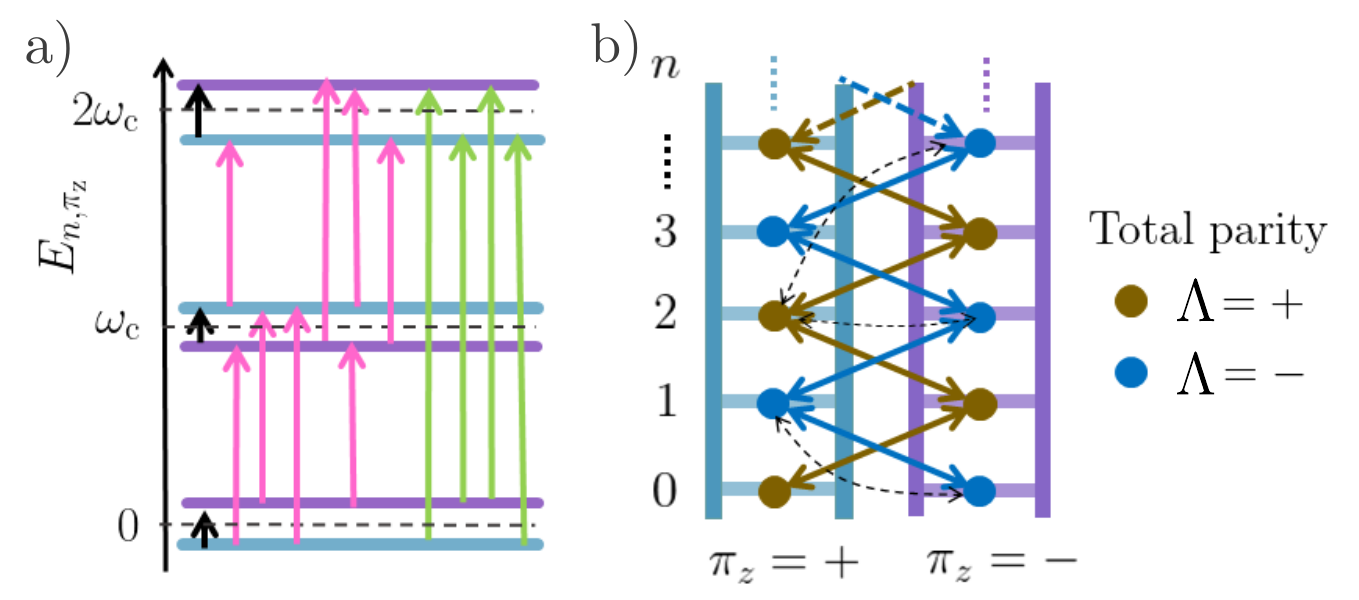}
\caption{a) Illustration of transitions, denoted by arrows, between the three lowest parity pairs of energy levels $E_{n,\pi_z}$. Transition arrows for Bohr frequencies $\omega$ clustered around $n\omega_\text c$ are colored black ($n=0$), pink ($n=1$) and green ($n=2$). b) Sketch of incoherent transitions (double sided arrows) between system states (filled dots) due to coupling to the thermal bath. Transitions at $\eta \rightarrow  \infty$ (thick, solid lines) change $\pi_z$ and move up or down one rung on the parity energy ladders, resulting in dynamics along two uncoupled state chains with conserved total parity $\Lambda=+$ (brown) and $-$ (blue). Finite $\eta$ leads to additional transitions (thin, dashed lines) between and within the two state chains (some examples shown).}
\label{fig:ugksl_ladder2}
\end{figure}

\textcolor{PS-color}{As is clear from Fig. \ref{fig:system} and Eq. (\ref{eq:energy}), the quasi-degenerate levels in the pDSC regime prevent a straightforward application of the secular approximation. Instead, we follow recent work \cite{pottsThermodynamicallyConsistentMarkovian2021a,trushechkinUnifiedGoriniKossakowskiLindbladSudarshanQuantum2021} and group the Bohr frequencies into clusters centered around frequencies $n\omega_\text c$, with a frequency spread of each cluster of the order of the energy level parity pair splitting, much smaller than the cluster separation $\omega_\text c$. This grouping, illustrated in Fig. \ref{fig:ugksl_ladder2}, allows the secular approximation to be safely applied for frequencies $\omega$ and $\omega^\prime$ in Eq. (\ref{BR}) belonging to different clusters. Further, assuming a temperature $k_\text BT \gg |E_{n,+}-E_{n,-}|$ for all relevant levels, we apply the singular coupling limit approximation \cite{pottsThermodynamicallyConsistentMarkovian2021a} and put $\gamma(\omega)=\gamma(n \omega_\text c)$, with $n = 0, \pm 1, \pm 2 \ldots,$ for all $\omega$ within the same cluster, centered at $n\omega_\text c$. These approximations reduce the Bloch-Redfield equation in Eq. (\ref{BR}) to an equation in Lindblad form, in the Schrödinger picture written as \cite{pottsThermodynamicallyConsistentMarkovian2021a, trushechkinUnifiedGoriniKossakowskiLindbladSudarshanQuantum2021}}
\begin{equation}
  \partial_t\hat \rho=-i[\hat H_D,\hat \rho]+ \sum_n \gamma(n\omega_c) \mathcal{D}[\hat A_n]\hat \rho
\label{Lindblad1}
\end{equation}
\textcolor{PS-color}{where $\mathcal{D}[\hat A]\hat \rho =\hat A \hat \rho \hat A^\dagger - \left\{\hat A^\dagger \hat A,\hat \rho\right\}/2$ and $\hat A_{n}=\sum_{\omega}\hat A(\omega)$, with the sum running over all frequencies $\omega$ belonging to the cluster around $n\omega_\text c$.}

This Lindblad equation (\ref{Lindblad1}), based on the properties of the spectrum in the pDSC limit, is still formulated in terms of the exact eigenstates of $\hat H_\text D$, preventing both an analytical treatment and a simple, compelling picture of the underlying physics. To provide such a picture we first consider the infinite coupling limit, $\eta \rightarrow \infty$. In this limit $\hat H_\text D=\omega_c\hat c^\dagger \hat c$, one HO per parity, with eigenstates $|n,\pi_z\rangle_0$. Only two hopping terms in Eq. (\ref{Lindblad1}), $\hat A_{-1}=\hat c^\dagger\hat \pi_x$ and $\hat A_{1}=\hat c \hat \pi_x$, are non-zero. The resulting equation is readily solved in the $|n,\pi_z\rangle_0$ basis, the probabilities $P_{n}^{\pi_z}=\rho_{nn}^{\pi_z \pi_z}\equiv {}_0\langle n,\pi_z| \hat \rho |n,\pi_z\rangle_0$ decouple from the coherencies and are determined by a rate equation 
\begin{align}
 \label{eq:rate}
 \frac{dP_n^{\pi_z}}{dt} =& \kappa (n_\text B(\omega_\text c)+1) \left[(n+1)P_{n+1}^{\bar \pi_z} - nP_n^{\pi_z}\right] \\
 \nonumber
 &+ \kappa n_\text B(\omega_\text c)\left[nP_{n-1}^{\bar \pi_z } - (n+1)P_n^{\pi_z}\right]   
\end{align}
\textcolor{PS-color}{where $\bar \pi_z=-\pi_z$. The dynamics is illustrated in Fig. \ref{fig:ugksl_ladder2}: The system undergoes incoherent transitions between states with opposite parity, up and down one rung on the parity energy ladders, due to the hopping terms $\hat c^\dagger \hat \pi_x$ and $\hat c \hat \pi_x$. This leads to the formation of two uncoupled chains of states with conserved {\it total} parity - both $\hat H_\text D$ and $\hat A_{\pm 1}$ commute with the total parity operator $\hat \Lambda=\hat \pi_z(-1)^{\hat c^\dagger \hat c}$. We find \cite{SM} that both the sum, $P_n^{+}+P_n^{-}$, and the staggered difference, $(-1)^n(P_n^{+}-P_n^{-})$, of the probabilities obey the standard rate equation for a cavity coupled to a bath and are subjected to sum rules $\sum_n(P_n^++P_n^-)=1$ (overall normalization) and $\sum_n (-1)^n(P_n^+-P_n^-) = \braket{\hat\Lambda}$, i.e. $\braket{\hat\Lambda}$ is a constant of motion. The steady solution is $(P_n^{\pi_z})^\text{ss}=(1-e^{-\beta \omega_\text c})e^{-n\beta \omega_\text c}(1+\pi_z(-1)^n\langle \hat \Lambda \rangle)/2$. }

To also include the corrections due to finite $\eta$, capturing the entire pDSC regime, the first step is to change the basis in Eq. (\ref{Lindblad1}) by performing a Schrieffer-Wolff (SW) transformation, making the Hamiltonian diagonal to first order in the perturbation. Formally, $\hat H_\text{SW}=e^{\hat S}\hat H_\text De^{-\hat S}=\hat H_0+\delta \hat H$ where the diagonal term $\hat H_0=\sum_{n,\pi_z}\epsilon_{n,\pi}|n,\pi_z\rangle_0{}_0\langle n,\pi_z|$, the small, second-order term $\delta \hat H=[\hat S,\hat{V}_D]/2$ where $\hat{V}_D$ are the off-diagonal components of $\hat{H}_D$ and $\hat S=-\hat S^\dagger$ with elements $S_{nm}^{\pi_z \pi_z}=\pi_z(\omega_\text q/2)D_{nm}^{\pi_z \pi_z}/(\epsilon_{n,\pi_z}-\epsilon_{m,\pi_z})$ for $n\neq m$, see endmatter B for details.

Neglecting the small $\delta \hat H$, the diagonal Hamiltonian $\hat H_0$ is the sum of $\omega_\text c\hat c^\dagger \hat c$, the unperturbed HO term, and a term describing the small parity splitting, with elements $\pi_z\omega_\text q D_{nn}/2$, see Eq. (\ref{eq:energy}). Since the two terms commute we can directly apply the scheme in \cite{pottsThermodynamicallyConsistentMarkovian2021a,trushechkinUnifiedGoriniKossakowskiLindbladSudarshanQuantum2021} and evaluate the dissipator in Eq. (\ref{Lindblad1}) taking the eigenstates in $\hat A_n$ to be the unperturbed $|n,\pi_z\rangle_{0}$. We also note that the SW-transformation modifies the system term in the system-bath Hamiltonian as $\hat A\rightarrow e^{\hat S}\hat A e^{-\hat S}\approx (\hat c^\dagger+\hat c)\hat \pi_x+[\hat S,(\hat c^\dagger+\hat c)\hat \pi_x]\equiv \hat A + \delta \hat A$, keeping terms up to first order in $\hat S$ (higher order terms give negligible corrections). Taken together, this results in an equation in the transformed SW-basis, in the Schrödinger picture given by
\begin{eqnarray}
\partial_t \hat{\rho}_\text{SW} &=& -i[\hat H_0,\hat{\rho}_\text{SW}] + \kappa (n_\text B(\omega_\text c)+1)\mathcal{D}[\hat \pi_x \desc] \hat \rho_\text{SW} \nonumber \\
&+& \kappa n_B(\omega_\text c)\mathcal{D}[\hat \pi_x \crec]\hat{\rho}_\text{SW}+ \delta \mathcal{D} \hat \rho_\text{SW}.   
     \label{eq:mainres}
\end{eqnarray}
\textcolor{PS-color}{The small term $\delta \mathcal{D} \hat \rho_\text{SW}=\kappa \sum_{k\geq 0}\{(n_\text B(k\omega_\text c)+1)\mathcal{D}[\delta \hat{A}_k] + n_\text B(k\omega_\text c)\mathcal{D}[\delta \hat{A}_k^\dagger]\}\hat\rho_\text{SW}$, with $\delta \hat A_k=\hat \pi_x \otimes \sum_n \langle n-k|_0{} \delta \hat A |n\rangle_0 |n-k\rangle_0{}_0\langle n|$, determine the long-time behavior of the dynamics, guaranteeing that the steady state solution is given by $\hat \rho_\text{SW}^\text{ss}=e^{-\beta \omega_\text c \hat c^\dagger \hat c}/\mbox{tr}(e^{-\beta \omega_\text c \hat c^\dagger \hat c})$, i.e., a Gibbs state with equal weight for the two parities. Note that in Eq. (\ref{eq:mainres}) the probabilities and coherencies in the unperturbed basis decouple. In terms of the solution $\hat \rho_\text{SW}(t)$ to Eq. (\ref{eq:mainres}), the expectation value of an observable $\hat O$ in the original frame, $\langle \hat O \rangle \equiv \mbox{tr}[\hat O \hat  \rho]=\mbox{tr}[\hat O_\text{SW} \hat  \rho_\text{SW}]$, where $\hat O_\text{SW}=e^{\hat S}\hat O e^{-\hat S}$, is thus given by }
\begin{equation}
\langle \hat O \rangle=\mbox{tr}\left[(\hat O+[\hat S,\hat O]+\frac{1}{2}[\hat S,[\hat S,\hat O]])\hat \rho_\text{SW}\right],
\label{eq:rotate}
\end{equation}    
where keeping terms up to second order in $\hat S$ is needed to correctly describe the effect of finite $\eta$. Eqs. (\ref{eq:mainres}) and (\ref{eq:rotate}) illustrate the additional benefit, besides guaranteeing a thermodynamically consistent framework, of moving to the SW frame: for short time scales (see discussion below), where the small term $\delta \mathcal{D} \hat \rho_\text{SW}$ can be neglected, Eq. (\ref{eq:mainres}) reduces to the rate equation (\ref{eq:rate}), exactly solvable and independent on $\eta$. The $\eta$-dependence of the observables enters the result via the transformation in Eq. (\ref{eq:rotate}), possible to perform analytically. Eq. (\ref{eq:mainres}) together with Eq. (\ref{eq:rotate}) thus provide a complete description of the dynamics in the pDSC regime and constitute the main result of the paper.

%% file: results_ps.tex
\textit{Illustration of ME}---We analyze Eqs. (\ref{eq:mainres}),(\ref{eq:rotate}) by benchmarking the time evolution and steady state results of the three expectation values $\braket{\crec \desc}\!(t)$,$\braket{\hat{\pi}_z}\!(t)$ and $\braket{\hat{\pi}_z \hat c}\!(t)$ against the BR result, Eq. (\ref{BR}). All numerical results were obtained using QuTip 5.2.3 \cite{qutip}.  
\begin{figure}[t]
    \centering
    \includegraphics[width=\linewidth]{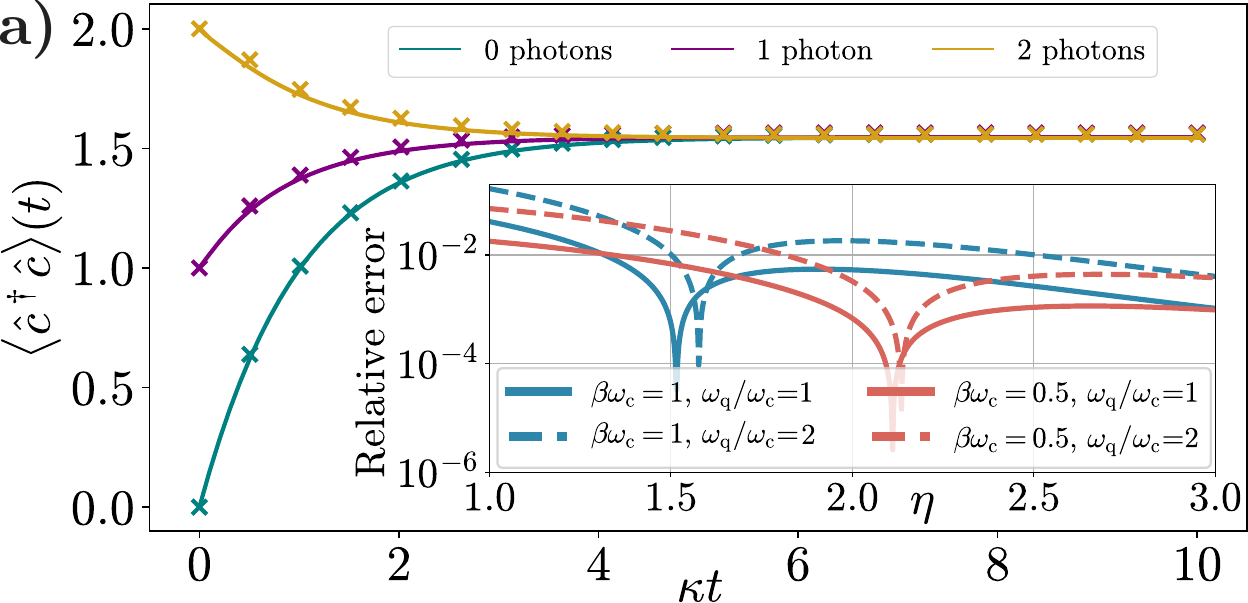}
    \includegraphics[width=\linewidth]{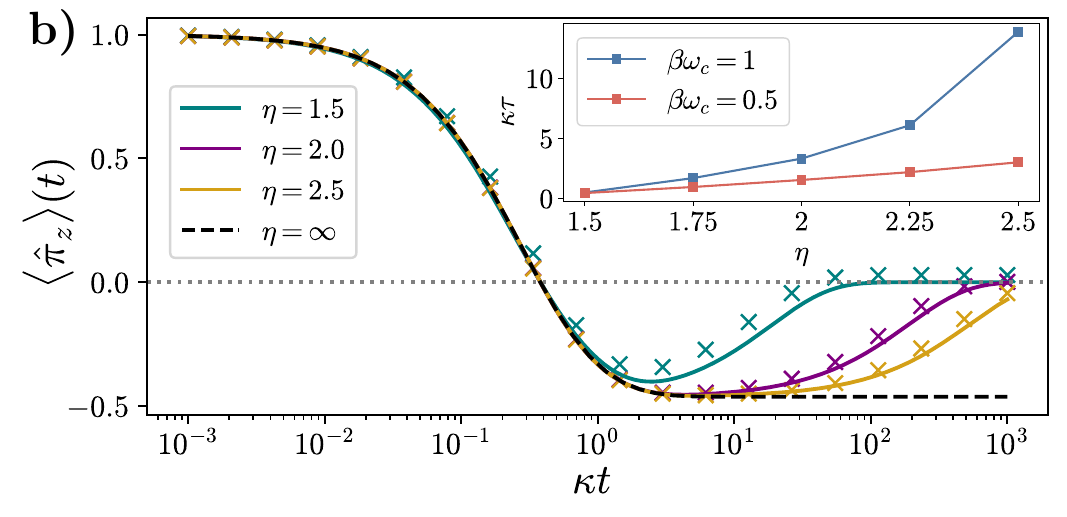}
    \includegraphics[width=\linewidth]{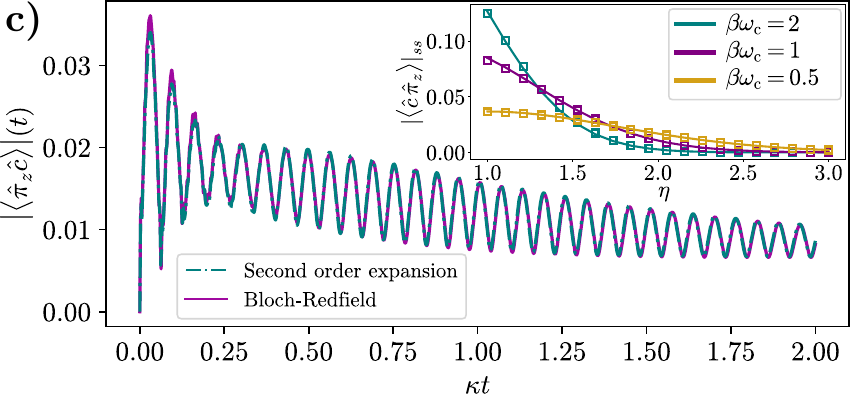}
    \caption{Benchmarking observables. In the main panels, the time evolution of a) occupancy $\langle \hat c^{\dagger} \hat c\rangle(t)$, b) parity $\langle \hat \pi_z\rangle (t)$ and c) magnitude of coherence $|\langle \hat c \hat \pi_z\rangle (t)|$ are shown. The results of Eqs. (\ref{eq:mainres}) and (\ref{eq:rotate}) (solid lines) and the Bloch-Redfield equation [crosses in a),b), purple in c)] are in very good agreement. For a), the initial, $t=0$, states $|n,+\rangle_0$ with the number of photons $n=0,1,2$. Here $\eta=2,\omega_\text q=\omega_\text c$ and $\beta \omega_\text c=0.5$. Inset: The relative error of the steady state occupations $|\langle \hat c^{\dagger} \hat c\rangle^\text{ss}-\langle \hat c^{\dagger} \hat c\rangle^\text{ss}_\text{BR}|/|\langle \hat c^{\dagger} \hat c\rangle^\text{ss}_\text{BR}|$ as a function of $\eta$ for different values of $\omega_\text q/\omega_\text c$ and $\beta \omega_\text c$. In b), the initial state $|1,+\rangle_0$ and $\eta=1.5,2$ and $2.5$, for $\omega_\text q=\omega_\text c$ and $\beta \omega_\text c=1$. The result without the term $\delta \mathcal{D} \hat \rho_\text{SW}$ in Eq. (\ref{eq:mainres}) shown as dashed. Inset: The cross-over time $\tau$ (see text) as a function of $\eta$ for different temperatures, with $\omega_\text q=\omega_\text c$. For c), taking $\delta H$ into account allows the transient coherent dynamics to be captured, initial state $|1,+\rangle_0$, $\eta=2$, $\omega_\text q=\omega_\text c$ and $\beta \omega_\text c=1$. Inset: Steady state results for Eqs. (\ref{eq:mainres}) and (\ref{eq:rotate}) (solid lines) and the Bloch-Redfield equation (squares) for different temperatures.}
    \label{fig:results}
\end{figure}
For the occupation $\braket{\crec \desc}\!(t)$, in Fig. \ref{fig:results}a), the agreement with the BR result in the pDSC regime is excellent, for different initial states, temperatures and ratios $\omega_\text q/\omega_\text c$. In all cases, the occupation approaches monotonically the steady state result $\braket{\crec \desc}_\text{ss}=n_\text B(\omega_\text c)+\delta n$, where the small correction can be written $\delta n=(\omega_\text q^2/2\omega_c^2)\exp[-4\eta^2\coth(\beta \omega_\text c/2)]\sum_{k=1}^{\infty}c_k$ with $c_k=\sinh(k\beta\omega_{\text{c}}/2)I_k[4\eta^2/\sinh(\beta \omega_\text c/2)]/k$ with $I_k(x)$ modified Bessel functions of the first kind. For $\eta \rightarrow \infty$, where the Rabi system can be described as two uncoupled, displaced oscillators, the occupation for $\beta \omega_\text c\gg 1$ decays as expected towards zero, as $\omega_\text q^2/(16\omega_\text c^2) \eta^{-2}$. Pushing the system parameters outside the pDSC regime, by increasing temperature or the ratio $\omega_\text q/\omega_\text c$ or decreasing $\eta$, the agreement with the BR equation deteriorates, as shown for the steady state in the inset of Fig. \ref{fig:results}a. 

The parity $\braket{\hat\pi_z}\!(t)$, shown in Fig. \ref{fig:results} b), is also in excellent agreement with the BR result in the pDSC regime. On timescales up to $\kappa t \sim 1$ the dynamics along the two total parity chains, well described by Eq. (\ref{eq:rate}), determine $\braket{\hat\pi_z}\!(t)$. For larger times, the small term $\delta \mathcal{D} \hat \rho_\text{SW}$ in Eq. (\ref{eq:mainres}) leads to a relaxation between the total parity state chains and hence equilibration of $\braket{\hat\pi_z}\!(t)$ towards $0$. The time $\tau$ for the cross-over to parity equilibration increases rapidly with increasing $\eta$ and also with decreasing $T$, as shown in the inset. For clarity, the result when omitting $\delta \mathcal{D} \hat \rho_\text{SW}$, i.e. the solution obtained with Eq. (\ref{eq:rate}), is shown with a dash-dotted line in the main panel. 

The quantity $\langle \hat c \hat \pi_z \rangle(t)$, shown in Fig. \ref{fig:results} c), captures the coherent oscillations of frequency $\approx\omega_\text c$ between the total parity state chains. To capture these transient oscillations a proper inclusion of $\delta \hat H$ into the coherent dynamics must be considered. Due to the smallness of $\delta \hat{H}$, the coherent dynamics can be calculated perturbatively from Eqs. \eqref{eq:rate} and \eqref{eq:rotate}, up to $t <\tau$ \cite{SM}. The amplitude of the oscillations, depending on $\eta, \beta\omega_\text{c}$ and $\omega_\text{q}/\omega_\text{c}$, decays towards a steady state value $(\omega_\text q/\omega_\text c)\eta \exp[-2\eta^2 \tanh (\beta \omega_\text c/2)]\tanh^2\left(\beta\omega_{\text{c}}/2\right)$ (see inset). Alternatively, adding $\delta \hat H$ directly to the Hamiltonian in Eq. (\ref{eq:mainres}) gives a good description of the coherent dynamics, including the correct long term behavior, but does not capture the amplitude of the first few coherent oscillations. We also note that including $\delta \hat{H}$ directly into \ref{eq:mainres} no longer guarantees formally that the steady state is the Gibbs state $\hat\rho_\text{SW}$. In all numerics performed we however find that $\hat\rho_\text{SW}$ is the steady state.

%% file: endmatter_ps.tex
\setcounter{equation}{0}
\renewcommand{\theequation}{A\arabic{equation}}
\textit{Appendix A --- Hamiltonian and electromagnetic gauge} In this letter we consider the quantum Rabi model - a two-level system coupled to a harmonic oscillator mode - typically an effective model of a larger interacting light-matter system where the matter sub-system has been truncated, or projected, to two levels. It has been shown in a number of works \cite{DeBernardis2018,distefanoResolutionGaugeAmbiguities2019,StokesNazir2023,sergi2024} that truncation or low-energy renormalization \cite{Lamberto2025} in the dipole gauge, in contrast to e.g. the Coulomb gauge, can accurately reproduce the spectrum of the untruncated light-matter system. For systems with large matter spectrum anharmonicity the accuracy can hold even at light-matter couplings well into the DSC regime, demonstrated e.g. for double well charge dipoles or a superconducting flux qubit coupled to a resonator \cite{DeBernardis2018}. This provides motivation for 
the quantum Rabi Hamiltonian $\hat H_\text R$ in Eq. (\ref{eq:Rabi}), expressed in the dipole gauge. Importantly, we do not make any claims about the gauge invariance of our truncated model, a topic that has been actively debated in recent years \cite{DeBernardis2018,StokesNazir2019,distefanoResolutionGaugeAmbiguities2019,Settineri2021,Savasta2021,StokesNazir2023,DiStefano2023Reply}, we merely claim that the predictions of our quantum Rabi model are accurate approximations to the untruncated theory, which is unambiguously gauge invariant.

For the open quantum Rabi system, we assume that the oscillator is linearly coupled to a continuum of external transverse electromagnetic modes, a standard description in cavity and circuit QED \cite{GardinerCollett1985,beaudoinDissipationUltrastrongCoupling2011}. Gauge consistency then requires that the dipole gauge should be applied also to the system-bath coupling \cite{Salmon2022}. The total Hamiltonian, describing the quantum Rabi system coupled to a thermal bath, is given by $\hat H_\text{tot}=\hat H_\text R +\hat H_\text B + \hat H_\text I$, where the bath Hamiltonian $\hat H_\text B=\sum_k \omega_k \hat b_k^\dagger \hat b_k$ and the system-bath coupling Hamiltonian
\begin{equation}
    \hat{H}_{\text{I}} = (\cre+\des+2\eta \hat\sigma_x)\otimes\sum_k (t_k \hat b^\dagger_k + t_k^* \hat b_k).
\end{equation}
Here $\hat{b}^\dagger_k/\hat{b}_k$ are the bosonic creation/annihilation operators of the bath modes with frequency $\omega_k$ and system-bath mode coupling strengths $t_k$. The system part, $\cre+\des+2\eta \sigma_x$, of $\hat H_\text I$, is obtained by performing \cite{Salmon2022} a unitary transformation, $\hat U=e^{\eta(\cre - \des)\hat \sigma_x}$, of the corresponding Coulomb gauge expression, $\cre+\des$, to the dipole gauge. We note that $\hat U$ is just the displacement transformation, i.e. $\hat D(\eta)$, discussed in the main text, explaining why the system part of the system-bath coupling Hamiltonian in the displaced frame is given by $\hat A=(\hat c^\dagger+\hat c)\hat \sigma_x$.

We stress that for a different type of resonator-bath coupling, or a bath different from a set of thermalized transverse electromagnetic modes, the coupling Hamiltonian $\hat{H}_{\text{I}}$ would be different \cite{StokesNazir2022}. An investigation of different $\hat{H}_{\text{I}}$ is however outside the scope of this work. 
\\
\\
\setcounter{equation}{0}
\renewcommand{\theequation}{B\arabic{equation}}
\textit{Appendix B --- Schrieffer-Wolff transformation} The starting point for the SW transformation is to write the Hamiltonian in the displaced basis, Eq. \eqref{eq:dispH}, as $\hat H_\text D=\hat H_0+\hat V$, where $\hat{H}_0=\sum_{n,\pi}(H_\text D)_{nn} |n,\pi_z\rangle_0{}_0\langle n,\pi_z|$, the sum of the diagonal elements of $\hat H_\text D$ in the unperturned basis $\ket{n,\pi_z}_0$, with $(H_\text D)_{nn}=\epsilon_{n,\pi}$ in Eq. (\eqref{eq:energy}). The term $\hat V$ contains the remaining, off-diagonal terms, and is given by
\begin{equation}
    \hat V =  \sum_{m\neq n,\pi_z=\pm}\pi_z\frac{\omega_\text{q}}{2}D_{nm}|n,\pi_z\rangle_0 ~_0\langle m,\pi_z|,
\end{equation}
where $D_{nm}=(-1)^{n+1}~_0\langle n|\hat D(-2\eta) |m\rangle_0$ for $n,m \geq 0$, with the matrix element of the displacement operator evaluated for the real parameter $\eta$
\begin{equation}
\begin{split}
&{}_0\langle n|\hat D(-2\eta) |m\rangle_0= \\
&e^{-2\eta^2}\times  
\begin{cases}
\sqrt{\dfrac{m!}{n!}}(-2\eta)^{n-m}
L_m^{(n-m)}(4\eta^2), & n\ge m,\\[2mm]
\sqrt{\dfrac{n!}{m!}}(2\eta)^{m-n}
L_n^{(m-n)}(4\eta^2), & m\ge n.
\end{cases}  
\end{split}
\label{Laguerre}
\end{equation}
Here $L_m^{n}(\alpha)$ is a generalized Laguerre polynomial. 

The operator, $\hat{S}$, implementing the SW transformation as $\hat H_\text{SW}=e^{\hat S}\hat H_\text De^{-\hat S}$, obeys $[\hat S,\hat H_0]=-\hat V$, which implies $\hat S^\dagger = -\hat S$ and guarantees a unitary transformation $e^{\hat S}$. In the unperturbed basis it has elements, diagonal in parity space, 
\begin{equation}
    S_{nm}^{\pi_z \pi_z}=\pi_z\frac{\omega_\text{q}}{2}\frac{D_{nm}}{\epsilon_{n,\pi_z}-\epsilon_{m,\pi_z}},\hspace{0.3 cm} n\neq m \geq 0
\end{equation}
as given in the main text. In the pDSC regime $S_{nm}^{\pi_z \pi_z}~\sim (\omega_\text q/\omega_\text c)D_{nm}/(n-m) \ll 1$, giving to leading non-zero order in $\hat S$ that $\hat H_\text{SW}= \hat H_0+\delta\hat H$, with $\delta\hat H=[\hat S, \hat{V}]/2$.

%% file: appendix_final_ps.tex
\section{System Hamiltonian spectrum and eigenstates}
Here we discuss the properties of the spectrum of the Rabi Hamiltonian in the pDSC regime, as well as the validity of the perturbative results. The Hamiltonian, $\hat H_\text D$, is given by Eq. (\ref{eq:dispH}). The eigenenergies for each parity $+,-$, calculated to second order in perturbation theory in $(\omega_\text q/2)(-1)^{\crec\desc}\hat{D}(-2\eta)$ - non-degenerate for each parity sector - are given by
\begin{equation}
E_{n,\pm}=n\omega_\text c\pm \frac{\omega_\text q}{2}D_{nn}+\frac{\omega_\text q^2}{4\omega_\text c}\sum_{m\neq n}\frac{D_{nm}D_{mn}}{n-m},
\label{eq:enpertpar}
\end{equation}
where the two first orders are denoted $\epsilon_{n,\pi_z}$ in Eq. (\ref{eq:energy}). Here where $D_{nm}=(-1)^{n+1}~_0\langle n|\hat D(-2\eta) |m\rangle_0$ for $n,m \geq 0$, with the matrix elements of the displacement operator given in Eq. (\ref{Laguerre}).

\begin{figure}[h]
    \centering
    \includegraphics[width=0.99\linewidth]{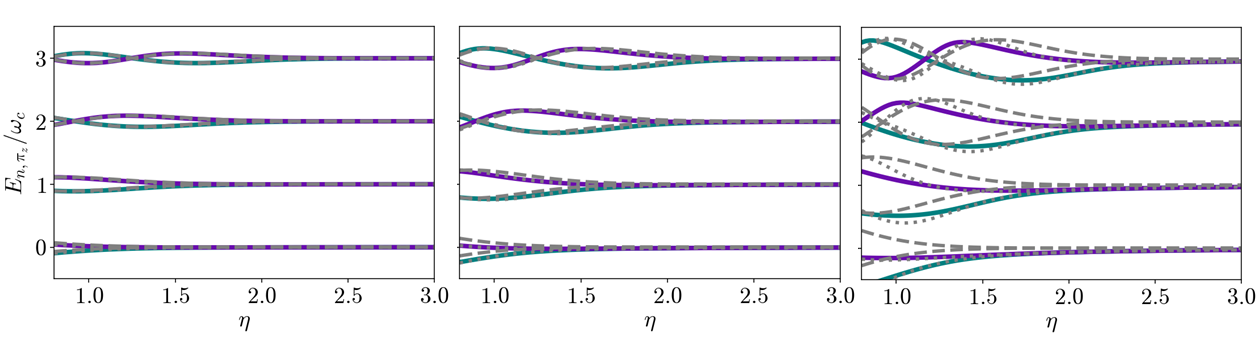}
    \caption{Exact and approximate spectra. The four lowest-in-energy parity pairs of the exact $\hat H_\text R$ eigenvalues are shown with thick, solid lines, teal (purple) for $+(-)$ parity, as a function of $\eta$. The ratio $\omega_\text q/\omega_\text c=1/2$ (left panel), $1$ (middle panel) and $2$ (right panel). The perturbative results to first (dashed line) and second (dotted line) order, from Eq. (\ref{eq:enpertpar}), are shown for comparison.}
    \label{fig:appedix_states1}
\end{figure}

To illustrate the range of validity for the perturbative result in Eq. (\ref{eq:energy}) in the pDSC regime we plot the exact result together with the first and second order expressions as a function of $\eta$ for different of $\omega_\text q/\omega_\text c$ in Fig. \ref{fig:appedix_states1}.
As is clear from the figure, while the first order perturbation result captures the spectrum in the pDSC regime well for $\omega_\text q/\omega_\text c \leq 1$, already for $\omega_\text q=2\omega_\text c$ the second order result is required to get overall energy correct. This justifies the claim in the main text that first order perturbation theory describes the pDSC for $\omega_\text q \lesssim \omega_\text c$.  

\section{Rate equation and population dynamics}
The rate equation for the occupation probabilities $P_n^+$ and $P_n^-$ in the displaced basis, Eq. (\ref{eq:rate}), can be solved exactly. The first step is to introduce 
sum, $P^\Sigma_n$, and staggered-difference, $P^\Delta_n$, populations, defined as 
\begin{equation}
     P^\Sigma_n=P_n^++P_n^-, \hspace{0.5 cm} P^\Delta_n = (-1)^n(P_n^+-P_n^-)
\end{equation}
By adding and subtracting the equations for $P_n^+$ and $P_n^-$ and expressing the sum and difference equations in terms of  $P^\Sigma_n$ and $P^\Delta_n$, we see that we get two identical, uncoupled equations, for $P^\Sigma_n$ and $P^\Delta_n$, as
\begin{equation}
    \dot P^X_n = \kappa (n_\text B(\omega_\text c)+1)\left[(n+1)P^X_{n+1}-nP^X_n\right]+\kappa n_\text B(\omega_\text c)\left[nP^X_{n-1}-(n+1)P^X_n\right], \hspace{0.3 cm} X=\Sigma, \Delta
    \label{eq:Xfunc}
\end{equation}
Eq. (\ref{eq:Xfunc}) is just the well known description for the dynamics of the number state probabilities of a cavity coupled to a thermal bath (See for instance \cite{Brange2019}). It follows from Eq. (\ref{eq:Xfunc}) that $\sum_n P^X_n$ is conserved, independent on time. We thus have $\sum_n P_n^\Sigma = 1$ (by normalization) and $\sum_n P_n^\Delta = \sum_n (-1)^n(P_n^+-P_n^-) = \braket{\hat\Lambda}$, where we recall that $\hat\Lambda=\hat{\pi}_z(-1)^{\hat c^\dagger \hat c}$ is the operator for total parity. The total parity $\braket{\hat\Lambda}$ is thus a constant of motion, as discussed in the main text.

To solve Eq. (\ref{eq:Xfunc}) we introduce the generating function $\mathcal{G}^X(z,t) = \sum_n z^n P^X_n(t)$
and inserting it into \eqref{eq:Xfunc} leads to a first-order PDE 
\begin{equation}
    \partial_t \mathcal{G}^X(z,t)=\kappa(1-z)\left[(1+n_\text B(\omega_\text c)-zn_\text B(\omega_\text c))\partial_z\mathcal{G}^X(z,t)-n_\text B(\omega_\text c)\mathcal{G}^X(z,t)\right].
\end{equation}
This equation can be solved by the methods of characteristics \cite{gardiner2009stochastic}. Presenting here no details of that calculation, we merely state that the solution is given by
\begin{equation}
    \mathcal{G}^X(z,t) = \frac{\mathcal{G}^X(z_0(z,t),0)}{1+n_\text B(\omega_\text c)(1-z)(1-e^{-\kappa t})}, \hspace{5pt} z_0(z,t) \;=\; 1-\frac{(1-z)\,\mathrm{e}^{-\kappa t}}
{1+n_\text B(\omega_\text c)\,(1-z)\big(1-\mathrm{e}^{-\kappa t}\big)},
\end{equation}
where we note that the initial populations, at $t=0$, enter the generating function via $\mathcal{G}^X(z_0(z,t),0)|_{t=0}=\mathcal{G}^X(z,0)=\sum_n z^n P^X_n(0)$. The time-dependent populations $P_n^X(t)$ are then obtained by successive derivatives of $\mathcal{G}^X (z,t)$ w.r.t. $z$, as
\begin{equation}
    P^X_n(t)= \frac{1}{n!}(\partial^n_z \mathcal{G}^X(z,t))\rvert_{z=0},
\end{equation}
Finally, the original probabilities are obtained by
\begin{equation}
    P_n^{\pi_z}(t) = \frac{1}{2}\left[P_n^\Sigma(t) + \pi_z (-1)^n P_n^\Delta(t)\right]
\end{equation}
In particular, in the long time, steady state limit we get the result in the main text, $(P_n^{\pi_z})^\text{ss}=(1-e^{-\beta \omega_\text c})e^{-n\beta \omega_\text c}(1+\pi_z (-1)^n\langle \hat \Lambda \rangle)/2$

\subsection{Example, parity relaxation}
To provide an illuminating example of the population dynamics, consider an initially unoccupied cavity with specified parity $\pi_z$, i.e. the initial state $\rho(0) = |0,\pi_z \rangle_0{}_0\langle 0,\pi_z|$ and hence $P_0^{\pi_z}(0)=1$ and all other probabilities zero at $t=0$. We then have
\begin{equation}
    \mathcal{G}^\Sigma(z,0) = 1, \hspace{0.3cm} \mathcal{G}^\Delta(z,0) = \pi_z \implies P^\Delta_n(t) = \pi_z P^\Sigma_n(t) = \frac{\pi_z  \tilde n(t)^n}{(1+\tilde n(t))^{n+1}},
\end{equation}
with $\tilde n (t) = n_\text B(\omega_\text c)\left(1-e^{-\kappa t}\right)$, giving population probabilities for a cavity with n photons and parity $\sigma$
\begin{equation}
    P_n^{\sigma}(t) = \frac{1}{1+\tilde n(t)}\left(\frac{\tilde n(t)}{1+ \tilde n(t)}\right)^n\left(\frac{1+\pi_z \sigma(-1)^n}{2}\right).
    \label{eq:probsigman}
\end{equation}
We then get the expectation values
\begin{equation}
    \braket{\hat\Lambda} = \pi_z, \hspace{5pt} \braket{\hat{\pi}_z}(t) = \sum_n (-1)^n P^\Delta_n(t) = \frac{\pi_z}{1+2\tilde n(t)}.
\end{equation}
i.e. the average parity decays towards the steady state value $\pi_z/(1+2n_\text B(\omega_\text c))$ on the time scale of $1/\kappa$.

\section{Coherent oscillations}
As pointed out in the main text, in the Lindblad equation (\ref{eq:mainres}) the diagonal and the off-diagonal, or coherent, terms in the unperturbed basis $|n,\pi_z\rangle$ decouple. As a consequence, the coherent oscillations of $\langle \hat c \hat \pi_z\rangle(t)$ shown in Fig. \ref{fig:results} c), calculated with the Bloch Redfield equation, are not captured by Eq. (\ref{eq:mainres}) and the subsequent (\ref{eq:rotate}). However, by adding the small, second order term $\delta \hat H$ to $\hat H_0$ in Eq. (\ref{eq:mainres}), the coherent oscillations are recovered, as shown in Fig. \ref{fig:results} c). Due to the smallness of the resulting coherent terms, they can be evaluated perturbatively and be expressed in terms of the diagonal, or probability, terms. For time scales shorter than $\tau$ [see Fig. \ref{fig:results} b)] the dissipator term $\delta \mathcal{D}\hat \rho_\text{SW}$ can be neglected and the probabilities are given by the rate equation (\ref{eq:rate}), with the exact solution given here in the section above.

To arrive at an expression for the coherences for times $t\lesssim \tau$ we start by writing the equation for the coherencies, the off-diagonal-in-photon-number density matrix elements $\rho_{n,m}^{\pi_z,\pi_z}\equiv \rho_{n,m}^{\pi_z}$ from Eq. (\ref{eq:mainres}), as
\begin{equation}
\begin{split}
    \Dot{\rho}_{n,m}^{\pi_z} &= -i\omega_\text{c}(n-m)\rho_{n,m}^{\pi_z} -i\delta\hat{H}_{n,m}\left(P_m^{\pi_z}-P_n^{\pi_z}\right) + \kappa(1+ n_\text B(\omega_\text c))\left(\sqrt{(n+1)(m+1)}\rho_{n+1,m+1}^{\bar \pi_z}-\frac{n+m}{2}\rho_{n,m}^{\pi_z}\right) \\ &+ \kappa n_\text B(\omega_\text c)\left(\sqrt{nm}\rho_{n-1,m-1}^{\bar \pi_z}-\frac{n+m+2}{2}\rho_{n,m}^{\pi_z}\right).
\end{split}
\label{eq:coherent}
\end{equation}
Here we recall that $\bar \pi_z=-\pi_z$ and note that we have taken $E_{n,\pi_z}=n\omega_c$, neglecting higher order corrections to the energies, at large times neglecting these corrections will lead to a noticeable phase drift and are therefore included in \ref{fig:results}c), here we neglect them for simplicity of presentation. From Eq. (\ref{eq:coherent}) it is clear that $\rho_{n,m}^{\pi_z}$ couple to the coherencies $\rho_{n\pm 1,m\pm 1}^{\bar \pi_z}$ as well as the probabilities $P_{n}^{\pi_z}$ and $P_{m}^{\pi_z}$. We can thus write the set of coupled linear equations for the coherencies compactly on a matrix form
\begin{equation}
    \Dot{\mathbf{c}}_{\pi_z,k} = (\boldsymbol{\Omega}_k + \mathbf{\Gamma}_k)\mathbf{c}_{\pi_z,k} + \mathbf{F}_{k}\mathbf{p}_{\pi_z,k},
    \label{eq:matrixeq}
\end{equation}
where we have introduced the coherence and probability vectors
\begin{equation}
    \mathbf{c}_{\pi_z,k}= \begin{pmatrix}  \rho_{0,k}^{\pi_z}, & \rho_{1,k+1}^{\bar \pi_z}, & \rho_{2,k+2}^{\pi_z}, & \dots  \end{pmatrix}^T, \hspace{0.5 cm}   \mathbf{p}_{\pi_z,k} = \begin{pmatrix}
        P_k^{\pi_z}-P_0^{\pi_z}, &P_{k+1}^{\bar \pi_z}-P_1^{\bar \pi_z}, & P_{k+2}^{\pi_z}-P_2^{\pi_z}, & \dots
    \end{pmatrix}^T,
\end{equation} 
the diagonal matrices $\mathbf{\Omega}_k$ and $\mathbf{F}_{k}$ with elements ($n=0,1,2,3,...$)
\begin{equation}
    (\mathbf{\Omega}_k)_{n,n}=ik\omega_\text c, \hspace{0.5 cm} (\mathbf{F}_{k})_{n,n}=-i\delta\hat{H}_{n,n+k}
\end{equation}
and the tri-diagonal matrix $\mathbf{\Gamma}_k$ with (non-zero) elements 
\begin{equation}
    \begin{split}
    (\mathbf{\Gamma}_k)_{n,n}&=-\frac{\kappa}{2}\left[n_\text B(\omega_\text c)(k+2n+2)+(n_\text B(\omega_\text c)+1)(k+2n)\right], \\
    (\mathbf{\Gamma}_k)_{n,n+1}&= \kappa( n_\text B(\omega_\text c)+1)\sqrt{(n+1)(k+n+1)}, \hspace{0.3cm} (\mathbf{\Gamma}_k)_{n+1,n}= \kappa n_\text B(\omega_\text c)\sqrt{(n+1)(k+n+1)}
    \end{split}
\end{equation}
The formal solution to Eq. (\ref{eq:matrixeq}) is
\begin{equation}
    \mathbf{c}_{\pi_z,k}(t) = e^{(\mathbf{\Gamma}_k+\boldsymbol{\Omega}_k)t}\mathbf{c}_{\pi_z,k}(0) + \int_{0}^t d\tau e^{(\mathbf{\Gamma}_k+\mathbf{\Omega}_k)(t-\tau)}\mathbf{F}_{k}\mathbf{p}_{\pi_z,k}(\tau).
    \label{eq:coherenceint}
\end{equation}
which is found to be in very good agreement with the numerics in Fig. \ref{fig:results} c). We also note that off-diagonal terms of $\mathbf{\Gamma}_k$ are needed to get the correct amplitude of the coherent oscillations. From Eq. (\ref{eq:coherenceint}) it is clear that the frequency of the oscillations is set by $\omega_\text c$, as pointed out in the main text. Moreover, the decay rate of the oscillations is governed by $\kappa$.

Formally, adding the term $\delta\hat H$ to the Hamiltonian in Eq. (\ref{eq:mainres}) implies that the steady state of the equation is no longer guaranteed to be the Gibbs state discussed in the main text. We however find from numerical investigations for a broad range of parameters in the pDSC that for times $t>\tau$ the dissipator term $\delta \mathcal{D}\hat \rho_\text{SW}$ dominates the dynamics, leading to a relaxation to the Gibbs state at long times.

\section{Steady-state expectation values}
Here we provide details for the calculation of the steady state expectation values of the occupation, parity and coherence, presented in the main text, as well as a generalization to other steady state observables. The starting point is Eq. (\ref{eq:rotate}), giving the expectation value of an operator $\hat O$ as sum over terms with commutators with respect to $\hat S$. To keep a general approach, we consider (normal ordered) operators $(\hat c^\dagger)^n\hat c^m$ and $\pi_z (\hat c^\dagger)^n\hat c^m$, with $n,m=0,1,2...$, which includes the three operators discussed in the main text. Focusing on the leading order corrections due to finite $\eta$ we can neglect the parity splitting of the energy levels in $\hat S$, i.e. take $E_{n,\pi_z}=n\omega_\text c$ and then write
\begin{equation}
    \hat{S} = \hat{\pi}_z \otimes \hat{G}, \hspace{0.5 cm}  {}_0\langle n|\hat G |m\rangle_0=\frac{1}{2}\frac{\omega_\text{q}}{\omega_\text{c}}\frac{D_{nm}}{n-m} \hspace{4pt} \text{for } n\neq m 
\end{equation}
Moreover, the steady state density matrix in the SW-frame is given by the Gibbs state
\begin{equation}
    \hat\rho_\text{SW}^\text{ss}=\boldsymbol{1}/2\otimes \hat \rho_\text{th}, \hspace{0.5 cm} {}_0\langle n|\hat \rho_\text{th} |m\rangle_0=(1-e^{-\beta\omega_\text c})e^{-n\beta \omega_\text c}\delta_{n,m}
\end{equation}
Inserting these expressions into Eq. (\ref{eq:rotate}) for the two observables we have to leading, non-zero order in $\hat G$
\begin{equation}
    \langle (\hat c^\dagger)^n\hat c^m \rangle^\text{ss} =\mbox{tr}\left[ (\hat c^\dagger)^n\hat c^m \hat\rho_\text{th}\right]+\frac{1}{2}\mbox{tr}\left[[\hat G,[\hat G, (c^\dagger)^n\hat c^m ]]\hat\rho_\text{th}\right], \hspace{0.3 cm}
     \langle \hat \pi_z(\hat c^\dagger)^n\hat c^m \rangle^\text{ss} =\mbox{tr}\left[ [\hat G,(\hat c^\dagger)^n\hat c^m]\hat\rho_\text{th}\right],
\end{equation}
where the traces run only over the photon spaces. Here we used that terms proportional to $\hat \pi_z$ have zero expectation value w.r.t. to $\hat\rho_\text{SW}^\text{ss}$. Inserting the expression for $\hat G$ and $\hat\rho_\text{th}$ and performing a set of tedious but straightforward reformulations we get for the three trace terms
\begin{equation}
    \begin{split}
        \mbox{tr}\left[ (\hat c^\dagger)^n\hat c^m \hat\rho_\text{th}\right]&=\delta_{nm}n!\left[n_\text{B}(\omega_\text c)\right]^n, \\    
        \mbox{tr}\left[ [\hat G,(\hat c^\dagger)^n\hat c^m]\hat\rho_\text{th}\right]&=-\frac{\omega_\text{q}}{\omega_\text{c} 2^{\frac{n+m}{2}+1}}
\frac{\sqrt{n!\,m!}}{|n-m|}
\frac{\mu^{-\frac{n+m}{2}-1}(\mu-1)^{\text{min}(n,m)}\left(\left(\mu+1\right)^{|n-m|}-\left(\mu-1\right)^{|n-m|}\right)}{(1+\mu)^{|n-m|/2}} \\
&\times e^{\eta^{2}\left(1-\frac{1}{\mu}\right)}\, 
\tilde D_{nm}, \hspace{4pt} \text{for } n \neq m \text{ and $0$ otherwise},\\
        \mbox{tr}\left[[\hat G,[\hat G, (c^\dagger)^n\hat c^m ]]\hat\rho_\text{th}\right]&=\frac{\omega_\text{q}^2}{2\omega_\text{c}^2(\mu+1)}
\sum_{\ell=0}^{\infty}\frac{\sqrt{(\ell+n)!(\ell+m)!}}{\ell!} \\
&\times \sum_{\substack{k\geq 0\\k\neq\ell+m,\;\ell+n}}\frac{\left[\left(r\right)^{\ell+m}+\left(r\right)^{\ell+n}-2\left(r\right)^k\right]D_{\ell+m,k}D_{k,\ell+n}}{(\ell+m-k)(k-\ell-n)},
        \end{split}
        \label{eq:sums}
\end{equation}
where we introduced the notation $\tilde \eta = \eta\sqrt{\frac{1+\mu}{2\mu}},\mu=\coth(\beta\omega_{\text{c}}/2)$, $r = \frac{\mu-1}{\mu+1}$ and $\tilde D_{nm}=(-1)^{n+1}~_0\langle n|\hat D(-2\tilde \eta) |m\rangle_0$ for $n,m \geq 0$.

\subsection{Occupation, parity and coherence}
Considering now the three cases discussed in the main text we have for the occupation, operator $\hat c^\dagger \hat c$, i.e. $n=m=1$, from Eq. (\ref{eq:sums}) that
\begin{equation}
     \langle \hat c^\dagger \hat c \rangle^\text{ss}=n_\text{B}(\omega_\text c)+\frac{\omega_\text{q}^2}{2\omega_\text{c}^2}
e^{-4\mu\eta^2}
\sum_{k=1}^{\infty}
\frac{1}{k}
\sinh\left(\frac{\beta \omega_{\text{c}} k}{2}\right)
I_k\!\left(4\eta^2\sqrt{\mu^2-1}\right).
\end{equation}
where $I_k(x)$ are the modified Bessel functions of the first kind and we used the Hardy-Hille formula. This is the result presented in the main text. In the limit of low temperature, $\beta \omega_\text c \gg 1$, the thermal occupation $n_\text B(\omega_\text c)$ is exponentially suppressed and the second term can be simplified, giving
\begin{equation}
    \langle \hat c^\dagger \hat c \rangle^\text{ss}= \frac{\omega_\text q^2}{4\omega_\text c^2}e^{-4\eta^2}\left[\mbox{Ei}(4\eta^2)-\ln(4\eta^2)-\gamma_\text E\right]  
\end{equation}
where $\mbox{Ei}(x)$ is an exponential integral function and $\gamma_E$ is the Euler gamma. Further, in the limit of $\eta \gg 1$ we have $\delta n \approx \omega_\text q^2/(16\omega_\text c^2) \eta^{-2}$, a power-law decay of the occupation with increasing coupling. For the parity, with operator $\hat \pi_z$, i.e. $n=m=0$, we have simply (as discussed in the main text)
\begin{equation}
    \langle \hat \pi_z \rangle^\text{ss}=0
\end{equation}
since $[\hat S,\hat \pi_z]=0$ and $\mbox{tr}[\hat \pi_z \hat \rho_\text{SW}^\text{ss}]=0$. For the coherence, operator $\hat c \hat \pi_z$, i.e. $n=0,m=1$, we have
\begin{equation}
     \langle \hat{\pi}_z \hat c \rangle^\text{ss}=\frac{\omega_\text q}{\omega_\text c} \eta \tanh^2\left(\frac{\beta\omega_\text{c}}{2}\right) \exp\left[-2\eta^2 \tanh \left(\frac{\beta \omega_\text c}{2}\right)\right],
    \label{app:cohtot}
\end{equation}
which is the result given in the main text.

\subsection{Smallness of discrepancy to Bloch Redfield result}
As shown in the main text, the steady state result in the pDSC agrees very well with the BR-result for all observables investigated. The smallness of the  discrepancy can be explained and quantified by noting the following: As pointed out in the main text, the steady state of the BR-equation, where we have discarded terms responsible for the Lamb-shift, is guaranteed \cite{Lee2022} to be a Gibbs state with respect to the full Hamiltonian, as
\begin{equation}
    \hat \rho^\text{ss}=\frac{e^{-\beta \hat H_\text D}}{\mbox{tr}(e^{-\beta \hat H_\text D})}
\end{equation}
We can then write the steady state expectation value for any observables $\hat O$ as
\begin{eqnarray}
    \langle \hat O \rangle_\text{BR}^\text{ss}&=&\mbox{tr}\left[\hat O \hat \rho^\text{ss}\right]=\frac{1}{\mbox{tr}(e^{-\beta \hat H_\text D})}\mbox{tr}\left[\hat O  e^{-\beta \hat H_\text D}\right]=\frac{1}{\mbox{tr}(e^{-\beta \hat H_\text D})}\mbox{tr}\left[e^{\hat S}\hat O e^{-\hat S} e^{-\beta e^{\hat S}\hat H_\text D e^{-\hat S}}\right] \nonumber \\
    &\approx& \frac{1}{\mbox{tr}(e^{-\beta \hat H_0})}\mbox{tr}\left[\left(\hat O +[\hat S,\hat O]+\frac{1}{2} [\hat S,[\hat S,\hat O]] \right)e^{-\beta \hat H_0}\right]\approx \langle \hat O \rangle^\text{ss}+\langle \delta \hat O\rangle^\text{ss}
\end{eqnarray}
Here we used that $e^{\hat S}\hat H_\text D e^{-\hat S}\approx \hat H_0+(1/2)[\hat S,\hat V]$ where we can write $\hat H_0=\omega_c \hat c^\dagger \hat c+\delta \hat H_0$, with $\delta \hat H_0=\hat \pi_z(\omega_\text q/2)\sum_n D_{nn} |n\rangle_0{}_0\langle n|$, writing the oscillator sector in the unperturbed basis. To leading order, expanding $e^{-\beta \hat H_0}\approx e^{-\beta \omega_\text c \hat c^\dagger \hat c}(1 - \beta \delta \hat H_0)$ and noting that $\langle \delta \hat H_0\rangle=0$, we can write the difference between the results from the BR-equation and our equation, as
\begin{equation}
    \langle \delta \hat O\rangle^\text{ss}=-\beta/2 \left(1-e^{-\beta \omega_\text c}\right)\mbox{tr}\left[\left(\hat O +[\hat S,\hat O]+\frac{1}{2} [\hat S,[\hat S,\hat O]] \right)\delta \hat H_0 e^{-\beta \omega_\text c\hat c^\dagger \hat c}\right]
\end{equation}
This trace can, similar to the ones in Eq. (\ref{eq:sums}) be further evaluated. Here, to provide a concrete example, we compare the deviation in parity we get to leading order in $\hat S$,
\begin{equation}
\langle \delta \hat \pi_z \rangle^\text{ss}= -\frac{\beta \omega_\text q}{2}\left(1-e^{-\beta \omega_\text c}\right)\sum_n D_{nn}e^{-\beta \omega_c n}=\frac{\beta \omega_q}{2}\tanh\left(\frac{\beta \omega_c}{2}\right)\exp\left[-2\eta^2\tanh\left(\frac{\beta \omega_c}{2}\right)\right].
\end{equation}